# PID Techniques: Alternatives to RICH Methods

J. Va'vra

SLAC National Accelerator Laboratory, CA 94025, USA

*Abstract* – In this review article[1] we discuss new updates on PID techniques, other than the Cherenkov method. In particular, we discuss recent efforts to develop high resolution timing, placing an emphasis on small scale test results.

## INTRODUCTION

Transition Radiation Detectors (TRD), dE/dx, and TOF detectors were reviewed previously at several RICH conferences [1,2]; therefore we will touch only briefly on these, and instead concentrate on new developments.

High luminosity LHC upgrades drive a new R&D effort, which plans to combine: (a) <u>timing</u> + position + front-end calorimetry, (b) <u>timing</u> to prevent event pile-up, and (c) <u>timing</u> + tracking. The new R&D stresses timing as a new important variable to resolve extremely large event multiplicities. Figure 1 shows the expected event pile-up per crossing as LHC luminosity increases towards $\sim 5\text{-}10 \times 10^{34}$ cm$^{-2}$ sec$^{-1}$. The association of the time measurement to the energy measurement is crucial for physics analysis, and requires time resolution of 20-30ps. On a longer time scale, one may combine timing and tracking; to do this, a new electronics developments are required. Another area of interest is pp-diffraction scattering at LHC. The aim is to use 2-proton-arm time difference to identify the collision point, and hence reduce pile-up background. A timing resolution of $\sim$10ps would locate $z_{vertex}$ to $\sigma \sim$2 mm, which is equivalent to $\sim$20x background rejection. LHC experiments give a high priority to these R&D efforts.

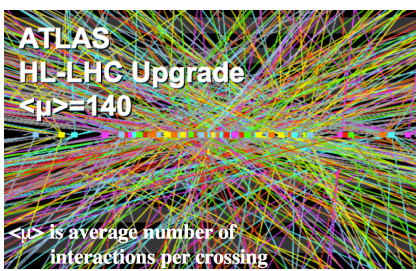

**Fig.1** This illustrates the large number of interactions per crossing after an LHC upgrade to a luminosity of $\sim 10^{35}$ cm$^{-2}$ sec$^{-1}$ (ATLAS collaboration).

large experiments. For example ALICE has improved TOF resolution by only a factor of two compared to MARK-III operating in 1980s. Yet, the ALICE experiment proved that by combining relatively modest TOF and dE/dx measurements, a good overall PID performance [5] can be achieved – see Fig.2. Given the long history of TOF development, the question is whether it can be improved significantly within a short period of time.

**Table 1:**

| Experiment | Method | Resolution | Comment |
|---|---|---|---|
| **MARK-II** (1975) | TOF (Scintillator) | 255 ps ($\sigma$) | Limited by poor $N_{pe}$ |
| **MARK-III** (1980) | TOF (Scintillator) | 140 ps ($\sigma$) | 3x larger $N_{pe}$ than MARK-II |
| **ALICE** (2016) | TOF (MRPC) | 85 ps ($\sigma$) | Limited by $t_0$ time, which depends on $N_{tracks}$ |
| **LBL TPC** (PEP4 - 1980) | dE/dx (TPC) | 2.4% ($\sigma_{dE/dx}$) | Pressure: 10 atm |
| **STAR TPC** (2003) | dE/dx (TPC) | 8% ($\sigma_{dE/dx}$) | Pressure: 1 atm |
| **ALICE** (2016) | dE/dx (TPC) | 7% ($\sigma_{dE/dx}$) | Pressure: 1 atm Pb-Pb collisions |
| **ATLAS** (2016) | TRD (TRT) | $\pi$-rejection factor ~100 | 1< p < 10 GeV/c |

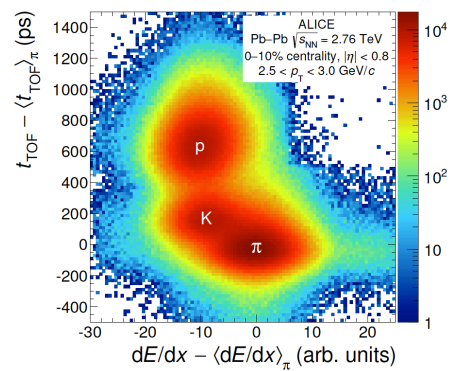

**Fig.2** (a) ALICE used a combination of both dE/dx and TOF, achieving a good PID separation up to ~3 GeV/c [5].

## 1. A BRIEF SUMMARY OF PID TECHNIQUES FROM EXISTING LARGE EXPERIMENTS

It took ~3 decades to develop dE/dx, TOF and TRD methods [1-4]. Table 1 shows a few examples of PID performance in

## 2. FAST TIMING AT A LEVEL OF 20-30 ps

A TOF technique seems simple, but there are many hidden effects which have to be overcome. There has been considerable progress with small single-pixel devices, however progress in very large systems has been slow. In the following I will discuss possible reasons.

One usually starts with a simple, naive formula in order to

---

This work supported by the Department of Energy, contract DEAC02-76SF00515.

[1] Invited talk at RICH 2016, September 8, 2016, Bled, Slovenia.

judge the timing resolution limit ($\sigma_{time}$):[2]

$$\sigma_{time} = \sigma_{noise}/(dS/dt)_{thereshold} \sim t_{rise\_time}/(S/N), \quad (1)$$

where S is the signal amplitude, N = $\sigma_{noise}$ is noise. For a typical MCP-PMT rise time of $t_{rise\ time} \sim$ 200 ps, one needs S/N ~10 to get into the ~20ps regime. For a slower Si-detector rise time of $t_{rise\ time} \sim$2 ns, one needs S/N ~100 to get into the same timing regime. However, this picture is still over-simplified, as there are many other contributions. First, we mention a few standard contributions due to electronics, chromatic effects, number of photoelectrons ($N_{pe}$), transit time spread (TTS), tracking, and $t_0$ time, which is experiment's start time (we neglect tracking effects as they depend on exact geometry):

$$(\sigma_{Total})^2 \sim [(\sigma_{Electronics})^2 + (\sigma_{Chromatic}/\sqrt{(N_{pe})})^2 + (\sigma_{TTS}/\sqrt{N_{pe}})^2 + (\sigma_{Track})^2 + (\sigma_{t0})^2] \quad (2)$$

The $t_0$ time can easily dominate. For example, the ALICE MRPC TOF detector timing is dominated by the $t_0$ time,[3] which strongly depends on the number of tracks in a given event, as shown in Fig.3a [6].

However, there are many additional effects,[4] which are impossible to express in a simple formula, and yet may contribute to the resolution at the 20-30ps level, if present:

a) Cross-talk can be a significant effect for very fast multi-pixel detectors, especially if different pixels have different pulse heights. Fig.3b shows the very complicated cross-talk observed in early versions of Planacon MCP-PMT [7]; this cross-talk was traced to an MCP backboard layout.

b) This cross-talk also created so called "coherent excitations", if five or more photons arrived at the same time on different pixels – see Fig.3c. This would upset the baseline, degrading the timing resolution of later-arriving photons. This cross-talk may exist in future fast multi-pixel Si detectors as well.

c) Another effect to worry about is the ion feedback in MCPs; ions are produced by electron bombardment of atoms, outgassed from the large MCP surface area. Fig.4a shows scope peaks corresponding to $H^+$, $H_2^+$ and $He^+$ ions; Fig.4b shows that the ion feedback rate gets worse at higher gain [8]. One of the worst tubes had 10 μm holes (TOF#1 in Fig.4b), which have larger total surface area. The operating voltage for these tubes was 2.4-2.5 kV to get single-photon sensitivity, i.e., close to the very rapid increase in ion feedback. Therefore the ion feedback rate should be checked, especially when it is planned to increase voltages to mitigate the gain loss at higher magnetic field values, or it is planned to use these tubes for a long time. One way to reduce ion feedback is to place an aluminum foil between top & bottom MCPs [9].[5]

d) The last effect we want to mention is a tail in the single-electron TTS spectrum[6], which may also affect timing resolution in detectors such as DIRC or TORCH. The origin of this tail is photoelectron recoil from the top MCP surface, as shown in Fig.5; the effect depends on the distance L between photocathode and top MCP surface. The Planacon MCP has L = 6 mm and a tail of ~0.6 ns, the Belle-II MCP has L = 2 mm and a tail of ~0.35 ns, and the stepped-face Photonis MCP has L = 0.85 mm giving almost no tail [10]. This effect is important in a multi single-photon, or a very high density track environment.

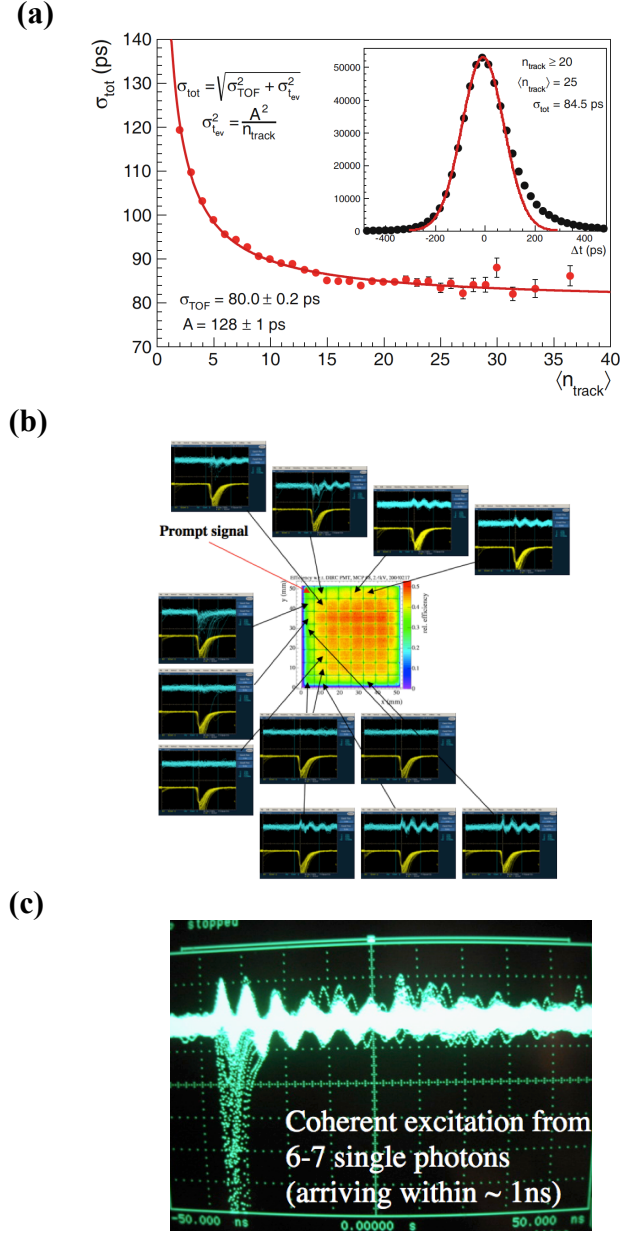

**Fig.3** (a) TOF resolution in the ALICE MRPC system. It is dominated by the $t_0$ start time resolution [6]. (b) The very complicated cross-talk observed in early versions of Planacon MCP-PMT by the FDIRC R&D effort [7]. (c) Coherent excitation in Planacon, driven by the cross-talk, if five or more photons arrive at the same time on various pixels [7].

---

[2] Well known formula in the communication engineering.
[3] The ALICE MRPC beam test has achieved $\sigma_{TOF} = t_{MRPC1}-t_{MRPC2}$ ~22 ps.
[4] In this paper we do not discuss software issues related to timing.
[5] A foil is the choice for the Belle-II TOP counter. At present, it is not known if this helps in the long run.
[6] TTS stands for the "Transit Time Spread" for single photoelectrons.

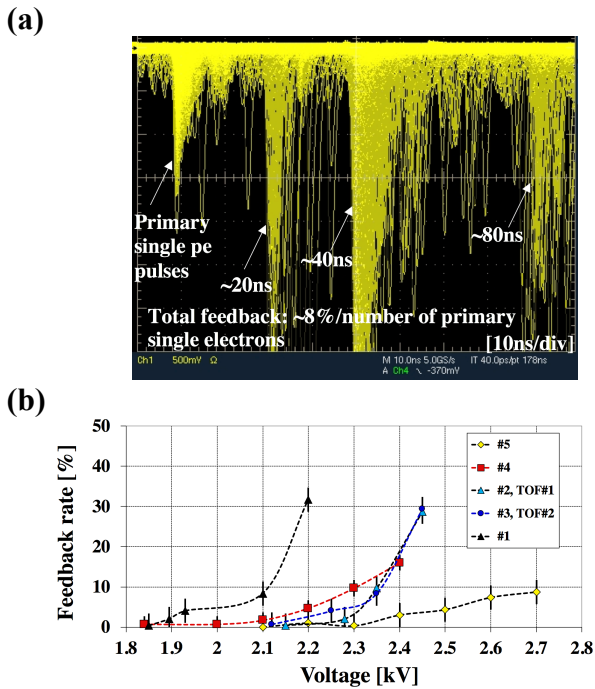

**Fig.4** (a) The ion feedback observed with the early versions of Burle Planacon MCP-PMTs. Peaks on a storage scope correspond to various ion drift times ($H^+$, $H_2^+$, $He^+$, etc.). (b) The ion feedback gets worse at higher gain. TOF#1 & TOF#2 tubes were used as TOF counters operating at ~2.2kV [8]. The feedback may get worse as tubes get older due to outgassing. Running MCPs at as low gain as possible helps to reduce the ion feedback.

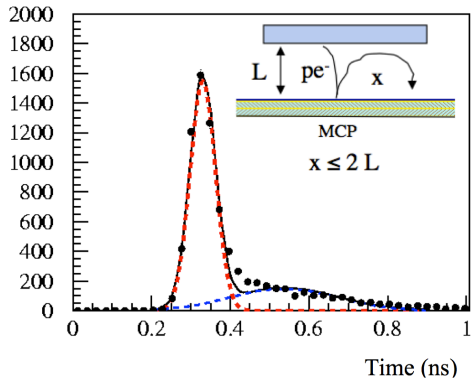

**Fig.5** A single photoelectron timing distribution for Planacon MCP, which has L = 6 mm. Tail, due to recoiling electrons, is dependent on the distance L between the photocathode and the top MCP surface [10].

## 2.1 Two examples of fast electronics

As this is not an electronics review, we will mention only two examples, which played a significant role in the early developments of fast detectors. The first one is the Ortec 9327 CFD + TAC analog electronics [11] – see Fig.6a.[7] The second one is the DRS4 waveform digitizing electronics, which is considered the leading performer in this particular field – see Fig.6b. S. Ritt measured <1 ps resolution for short delays [12].

---

[7] According to Jeff Peck, the Ortec engineer who designed this electronics, the 9327 CFD can reach ~2 ps resolution if one avoids using the TAC and chooses a good pulser.

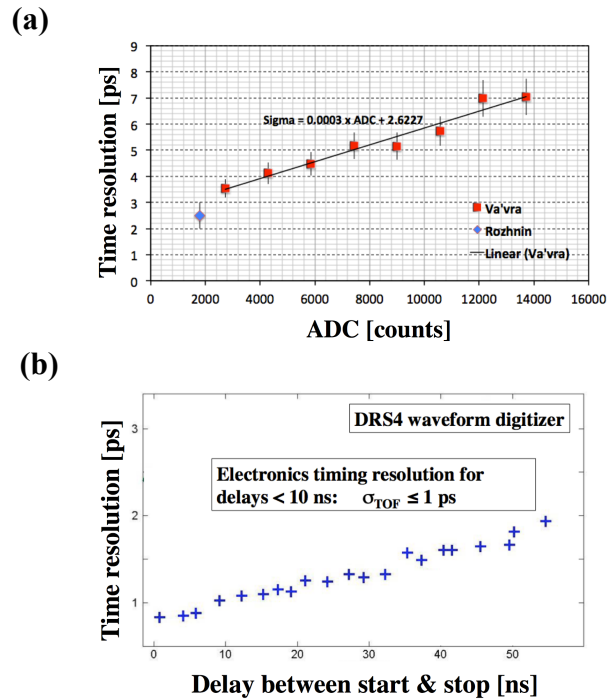

**Fig.6** (a) Electronics resolution obtained with Ortec 9327 CFD+TAC analog electronics[11]. (b) DRS4 timing resolution [12].

## 2.2 MCP-PMT tests

Figure 7a shows a timing resolution of $\sigma_{TOF}$ ~14 ps (TOF = $t_{MCP1}$ - $t_{MCP2}$) [13], operating at gain of ~$10^6$, no amplifier (this may be unrealistic in real experiments)[8], ~80 pe's per 120 GeV protons, and with the Ortec 9327CFD, TAC566, and ADC114 electronics. Two single-pixel Photek-240 MCPs were separated by 7.12 meters. They had 6 μm holes, and a 40 mm diameter, 8 mm-thick Fused Silica window.

In a separate test [13], Fermilab people improved the timing resolution to $\sigma_{TOF}$ ~9.6 ps (see Fig7b) with two back-to-back MCPs in one common RF shielded box, running with the DRS4 digitizer, and with no amplifier. Fig.7b also shows the old Nagoya result[6] of $\sigma_{TOF}$ ~8.8 ps [14], still the best of any detector. The Nagoya test had two back-to-back R3809U-59-11 MCPs with 6 μm holes, operated at a very high gain of ~$2\times10^6$, with quartz radiator length of 10 mm (plus 3 mm MCP window), and with no amplifier.[9] Recently, this test was repeated at CERN with the same R3809U MCPs and LeCroy 2.5 GHz scope [15]. They obtained $\sigma_{TOF}$ ~5.7 ps. **This result is the best timing result of any detector mentioned in this paper.**

However, all three tests used single-pixel special MCP detectors, did not use amplifiers, and operated at a very high gain, which is not realistic in real experiments. A more realistic test [16], operating at a modest gain of ~$10^5$, and with an amplifier, achieved $\sigma_{TOF}$ ~20 ps. The test had only a small number of photoelectrons (Npe ~23); with Npe ~80 pe's it could have achieved $\sigma_{TOF}$ ~15 ps.

---

[8] See earlier comments about ion feedback.
[9] A Nagoya paper [14] presented the resolution of a single MCP to be $\sigma_{Single\_MCP}$ ~6.2 ps (private communication with K. Inami). In this paper we present results as: $\sigma_{TOF} = \sigma_{MCP1} - \sigma_{MCP2}$, and try to be consistent throughout.

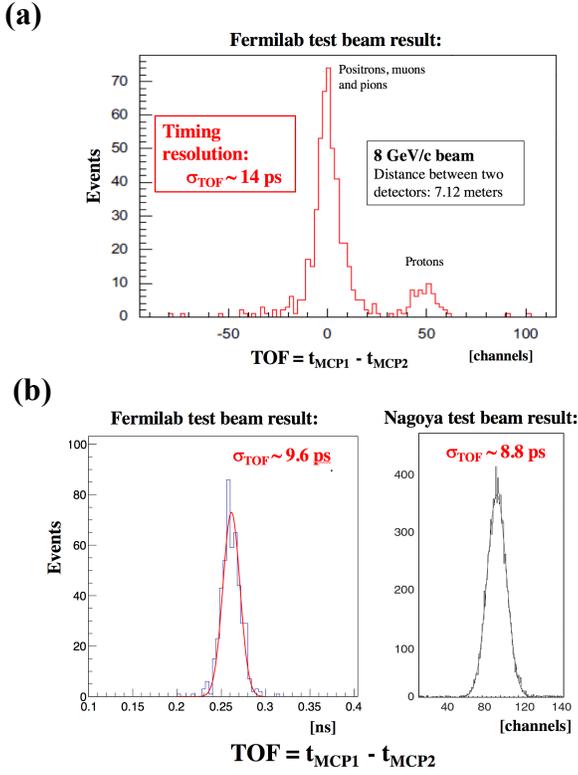

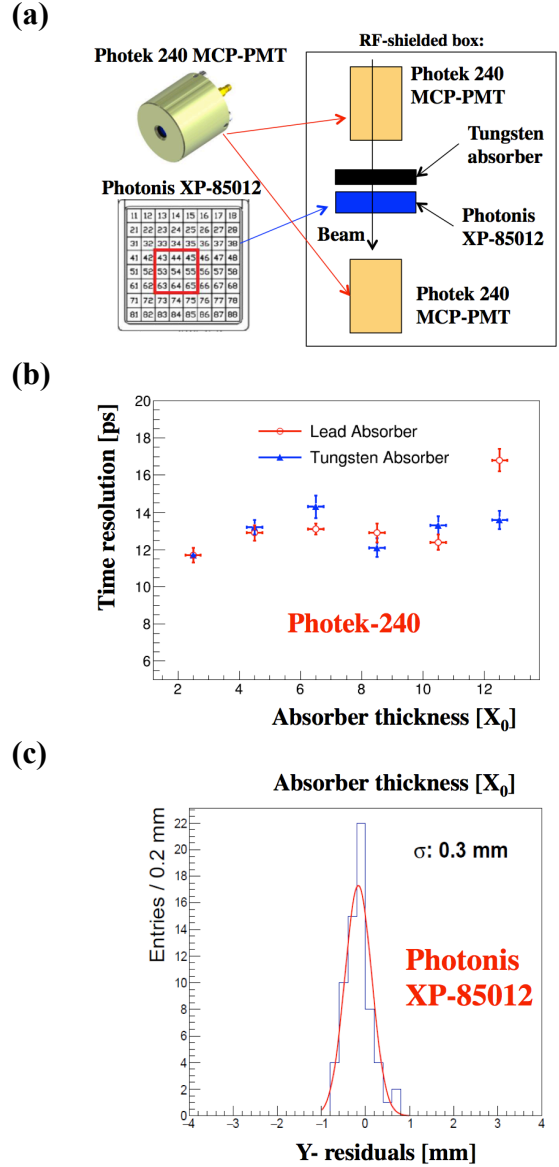

## 2.3 MCP-PMT tests for Calorimetry

MCP-PMTs can be used for fast calorimetry, as demonstrated in the Fermilab 8 GeV electron test beam [19]. Figure 9a shows the setup. Photek tubes were used for timing with the DRS4 electronics, and the Photonis XP-85012 to measure shower position. Figures 9b and 9c show a timing resolution of $\sigma_{TOF}$ ~15 ps (TOF = $t_{MCP1} - t_{MCP2}$) and a position resolution with $\sigma$ ~0.3 mm for absorber thickness of $6X_0$.

**Fig.7** (a) Fermilab TOF beam test with two Photek-240 MCPs, separated by 7.12 meters [13]. (b) Comparison of the two best timing results from any detector so far: one from the second Fermilab beam test with two back-to-back Photek-240 MCPs [13], and one from the old Nagoya test with two back-to-back R3809U-59-11 MCPs [14], both plotted in the same variable.

Figure 8 shows the LQbar detector, proposed by A. Brandt, and intended for the LHC pp-diffraction scattering experiment. It uses a mini-Planacon MCP-PMT with 10 μm pores, operating with 10x amplifiers. Preliminary beam test results show a resolution $\sigma$ ~40-45 ps per single bar, and $\sigma$ ~35 ps per group of 4 bars [17].[10] Based on laser tests [18], they found that running at a lower MCP gain of ~$10^5$ preserves good timing, if the number of photoelecteons is ≥10; low gain is essential for handling high rates. Use of optical grease did not help the resolution, as it reduces the number of photoelectrons.

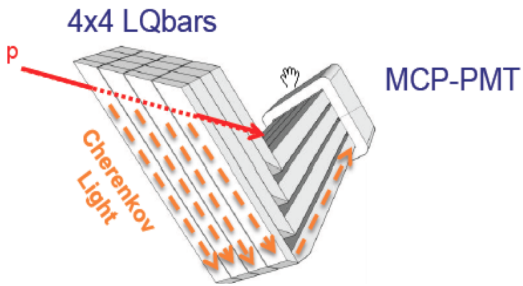

**Fig.8** LQbar MCP-PMT detector proposed for the pp-diffraction scattering experiment at the LHC [17].

**Fig.9** (a) Fermilab experimental setup. (b) Calorimeter timing resolution of $\sigma_{TOF}$ ~15 ps. (c) Position resolution of $\sigma$ ~0.3 mm, both at shower max [19].

The last comment is about the status of large size MCP-PMTs being developed by the LAPPD group. At the moment, the Incom Co. is trying to get preproduction going. None of the 20 cm x 20 cm MCPs has yet been tested in a beam; only laser tests were done while a detector was in the production line. The plan is to put 4 LAPPD detectors into the test beam soon [20]. There is a parallel R&D effort in Argonne trying to develop smaller pixilated MCPs [21].

---

[10] Results do not seem to scale correctly, possibly because of the MCP cross-talk effects mentioned earlier.

## 3. TIMING WITH SILICON

Avalanche diodes represent an old Si-based technology [22]. Recently these devices were considered for a high resolution timing by several groups. In the early days, people thought that these devices must be operated at high gain to obtain good timing resolution. This is still basically true, but as we will see, good timing resolution can be achieved at lower gain, provided that the S/N ratio is sufficiently high.

In this section we will deal mostly with single-pixel-detector test results, which means that results tend to be more optimistic.

Specifically, we will discuss recent results from these types of detectors: (a) Zero gain PIN diode; (b) High gain HAPDs and SiPMTs; (c) Medium gain Avalanche Diodes (MGAD); and (d) Low gain Avalanche Diodes (AD) = LGAD.[11]

### 3.1 Zero gain PIN diodes

How can a PIN diode with no gain be used for timing? Figure 10 shows that it is possible if one adds a $6X_0$ tungsten absorber in front of it, and uses an electron beam. A Fermilab group [19] achieved $\sigma_{TOF} \sim 23$ ps (TOF = $t_{MCP} - t_{PIN}$) in a 32 GeV electron beam. This result may be significant for the development of future fast calorimeters.

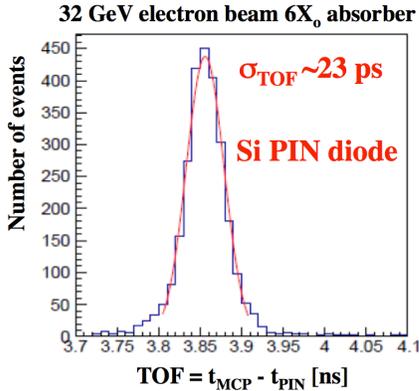

**Fig. 10** Timing resolution between PIN diode and Photek-240 MCP in a 32 GeV electron beam; the PIN diode had a $6X_0$ tungsten absorber in front of it [19].

### 3.2 High Gain HPDs and SiPMTs

The HPD is a ~15-year-old technology [23]. It is a very high gain device, where a larger portion of the gain comes from an energetic electron impact into a Si AD (~1600x), and a smaller fraction from the AD (~110x) itself, resulting in a good S/N ratio. A laser test with a single-pixel AD device achieved a single photoelectron resolution of $\sigma_{TTS} \sim 9$ ps for a laser spot size of 1 mm, and a Bialkali photocathode. Subsequent tests with a multi-pixel AD (see Fig.11) measured a transit time spread $\sigma_{TTS} \sim 90$-$100$ ps with pixel sizes of 2 mm x 2 mm [24].

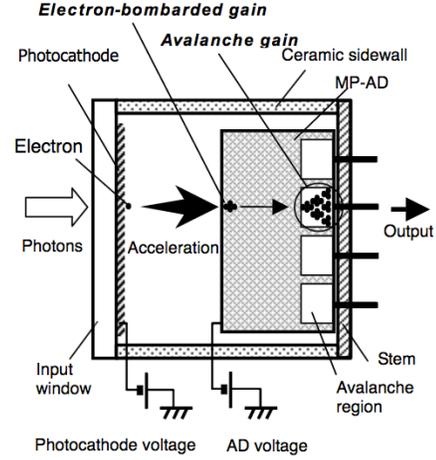

**Fig. 11** Multi-pixel HPD achieved $\sigma_{TTS} \sim 90$-$100$ ps [24].

The SiPMT is another high gain device. Figure 12 shows a SiPMT timing resolution of $\sigma_{TOF} \sim 16.3$ ps (TOF = $t_{MCP}$ - $t_{SiPMT}$), and $\sigma_{SiPMT} \sim 14.5$ ps, after the MCP contribution is substracted; this was measured in a 120 GeV proton beam [24]. SiPMT used a fused silica 30 mm x 3 mm$^2$ radiator, and was coupled to a fast amplifier. The Photek-240 MCP (6 μm holes), used as a stop, had no amplifier and had a 8 mm-thick fused silica window. The readout used the DRS4 digitizer.

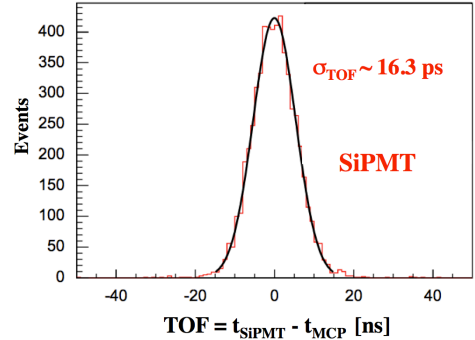

**Fig. 12** Timing resolution between an SiPMT and a Photek-240 MCP-PMT, as obtained in a 120GeV Fermilab proton test beam [25].

Figure 13 shows an interesting application of SiPMTs in the Panda TOF detector. It consists of small scintillator tiles (30 x 30 x 5 mm$^3$) made of fast EJ-228 scintillator. The entire system will have 5760 tiles covering an area of ~5.2m$^2$. A test in a 2.7 GeV/c proton beam achieved $\sigma_{TOF} \sim 82$ ps with analog SiPMTs[12], and $\sigma_{TOF} \sim 32$ ps with Philips digital SiPMTs.[13] The goal of the experiment is to reach $\sigma_{TOF} < 75$ ps [26]. Although not chosen at the end, the Philips digital photon counter concept[13] may indicate the future of Si-based multi-pixel detectors which integrate sensor, front end and digital electronics in the same package.

---

[11] In this paper we do not use names such as "hyper-fast" or "ultra-fast" silicon.

[12] Ketek SiPMT: http://www.ketek.net/products/sipm/
[13] Philips digital SiPMT: http://www.digitalphotoncounting.com/

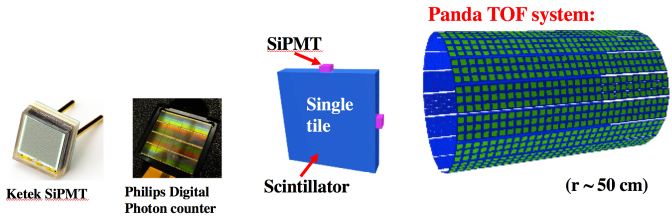

**Fig. 13** The Panda TOF system, based on SiPMTs and fast scintillator tiles [26].

### 3.3 Medium Gain Avalanche Diodes

Medium gain avalanche diodes (MGAD) are based on so called "Deep depleted" avalanche diode technology, which allows a gain of 200-500x. They were developed by the RMD Co. (Radiation Monitoring Devices Inc.), and have been around for more than 10 years [27]. This is a radiation-hard technology, tested up to ~$10^{15}$ neutrons/cm$^2$. Recent devices have a large sensitive area of ~1 cm x 1cm [28]. The amplification, located deep inside the bulk, requires a large electric field of ~15 kV/cm with a voltage of ~1.75 kV. Figure 14a shows the device, its cross-section profile, and a final arrangement as a test detector. Figure 14b shows its rise time of ~1.2 ns, compared to a ~200 ps MCP-PMT risetime. This means that the AD's S/N ratio needs to be ~5x better to get comparable resolution - see Eq. (1). Figure 14c shows that the group measured an excellent timing resolution of $\sigma_{TOF}$ ~18 ps (TOF = $t_{AD}$ - $t_{MCP}$) in the CERN muon test beam. The illuminated area was ~8 mm x 8 mm during the test. The electronics used a fast amplifier (made by M. Newcomer at al., or Wenteq with 50 dB gain) coupled to a SAMPIC digitizer (6.4 GSa/sec, 156 ps/step) [29]. Timing resolution obtained with a laser[14], for a MIP-equivalent charge, was also excellent: $\sigma_{TOF}$ ~12 ps (TOF = $t_{Trigger}$ - $t_{AD}$).

Many questions remain to be answered. For example, the impact of edge effects, or of cross-talk in multi-pixel devices.

Proponents believe that this device would be suitable for the future LHC pp-diffraction scattering application, and for a pre-shower detector in a fast calorimeter. Higher gain helps to achieve better risetime and the required S/N ratio.

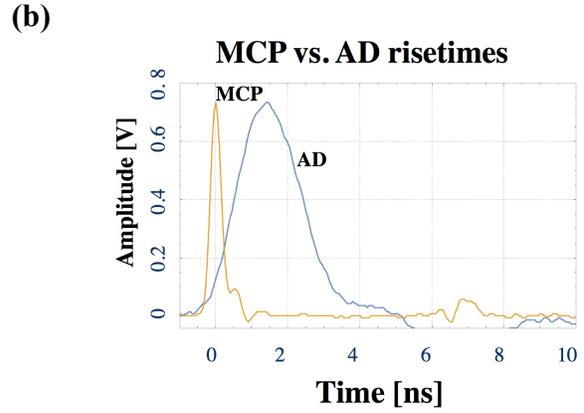

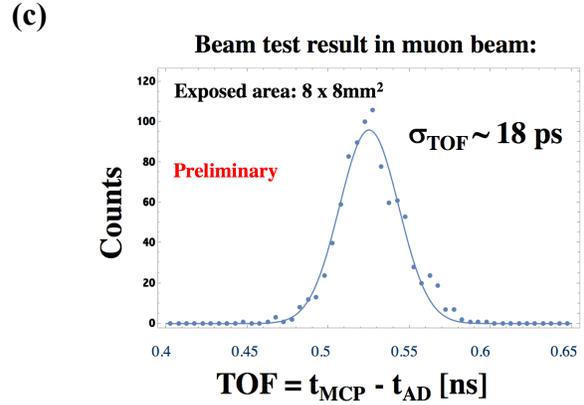

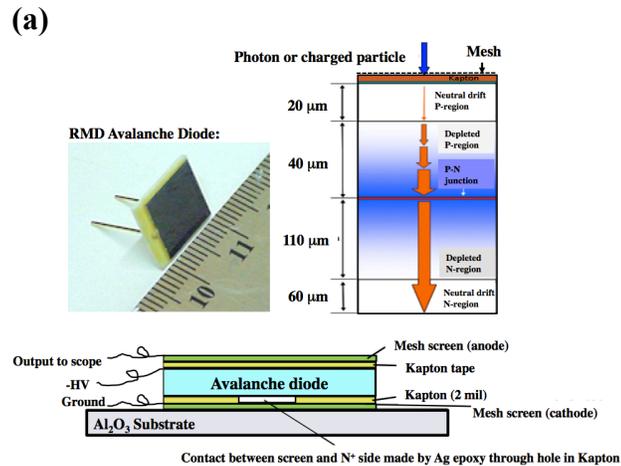

**Fig. 14** (a) The RMD avalanche diode of ~8 mm x 8 mm active area. (b) Risetime of RMD avalanche compared to that of MCP-PMT. (c) Test result in a muon test beam [28].

### 3.4 Low Gain Avalanche Diodes

Low Gain Avalanche Diodes (LGAD) are based on so called "Reach through thin AD technology," which can get a gain of up to 15-30x for voltages of 200-240 Volts. Figure 15a shows the device made by the Advanced Photonics Co.: ~1.3 mm x 1.3 mm active area, and 50 μm thickness. Figure 15b shows a timing resolution obtained with the CERN 180 GeV pion beam: $\sigma_{TOF}$ ~49 ps at 200V, and $\sigma_{TOF}$ ~36 ps at 240 V (TOF = $t_{LGAD1}$ - $t_{LGAD2}$) - see Fig.15b [30]. The electronics in this test used a fast amplifier Cividex 100x (1-2MHz BW), and a CFD coupled to the LeCroy scope (20 GSa/sec, 50 ps/bin). Authors define $\sigma(1) = (\sigma_{LGAD1-LGAD2})/\sqrt{2}$, and find that it scales as $\sigma(N) = \sigma(1) N^{-0.5}$, where N is the number of LGAD layers; for example, the time resolution of three LGAD layers: $\sigma(3)$ ~20 ps at 200 V, and $\sigma(3)$ ~15 ps at 240 V. Figure 15c shows a table sumarizing the results.

Proponents believe that this device will be able to combine timing with tracking, and so can be used for front-end calorimtery and in pp-diffraction scattering. However, it should be pointed out that there is no multi-channel ASIC available yet, which would offer a fast preamp. In addition, many detailed questions need to be answered, such as edge effects outside cell amplification boundaries, cross-talk among cells, sensitivity to breakdowns, etc.

---

[14] VCSEL laser operating at a 980 nm wavelength.

(a)

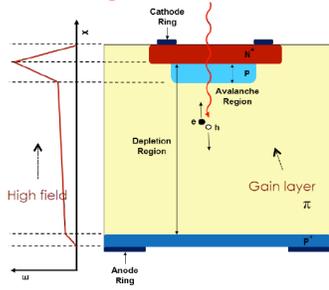

(b)

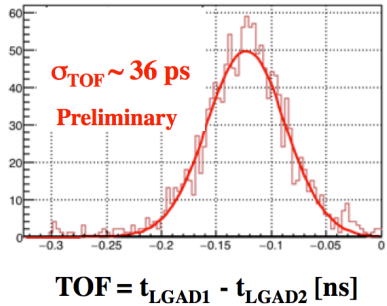

(c)

σ(N) = f(number of LGADs):

| Timing Resolution [ps] | | |
|---|---|---|
| Vbias [V] | 200V | 240V |
| N=1 : | 34.6 | 25.6 |
| N=2 : | 23.9 | 18.0 |
| N=3 : | 19.7 | 14.8 |

**Fig. 15** (a) Principle of Low Gain Avalanche Diodes (LGAD) with ~1.3 mm x 1.3 mm active area, and 50 μm thickness. (b) Test beam TOF result for two LGADs at 240 V [30]. (c) Test beam results as a function of number of LGAD layers for 200 and 240 V [30].

## 4. TIMING WITH DIAMOND

Good timing resolution can be achieved also with Ultra-pure mono-crystaline sCVD diamond detectors. They are fast, have low noise, and are radiation hard.

The TOTEM group [31] has tested a device with a pixel size of 4.5 mm x 4.5 mm, and 500 μm thickness. A MIP signal creates a very-low-signal of ~$1.2 \times 10^4$ electrons, and therefore very-low noise electronics must be used. Good results were obtained with a low noise charge amplifier with a ~10 ns shaper (a 2 GHz BW amplifier gave worse resolution). The detector had a high rate capability of up to ~3 MHz/cm$^2$. Figure 16 shows measured timing resolution of $\sigma_{TOF}$ ~90 ps (TOF = $t_{Diamond\_1}$ - $t_{Diamond\_2}$), obtained with the Agilent scope (2.5 GHz, 20 GSa/s).

Proponents hope that this device could be used for the future LHC pp-diffraction scattering application if they use several detector layers. This concept can also be used as a pre-shower detector for fast-timing calorimeters.

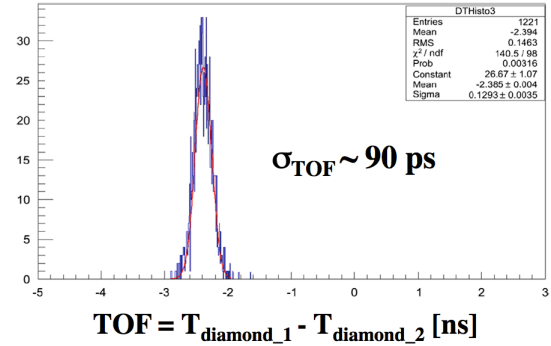

**Fig. 16** A diamond detector with a pixel size of 4.5 mm x 4.5 mm, and 0.5 mm thickness achieved $\sigma_{TOF}$ ~ 90 ps in a test beam [31].

## 5. TIMING WITH GASEOUS DETECTORS

It is suprising, but even gaseous detectors may have a chance of achieving good timing. Critical parameters are the number of photoelectrons contributing to a signal's leading edge, short drift distance, and a gas with a small longitudinal diffusion. The technology of choice is the Micromegas detector [32], coupled to a very short drift of ~200 μm, as shown on Figs.17a and 17b. Figure 17c shows the calculated photo-electron timing spread in a 200 μm drift length as a function of gain in this section. It can be seen that to keep the longitudinal diffusion below $\sigma_L$ ~300 μm, a high field $E_{drift}$ >10 kV/cm is needed where a gain is larger than 10.

The prototype had a 3 mm-thick $MgF_2$ window coated with a ~18 nm-thick CsI photocathode, pixel size of ~1cm$^2$, and used flowing Ne-$CF_4$-$C_2H_6$ gas. So far they have reached 8-10 photoelectrons contributing to the leading edge (rise time of ~2 ns). There is a long pulse tail, which does not contribute to timing, however, it may affect later-arriving pulses. Figure 17d shows a measured resolution of $\sigma_{TOF}$ ~35 ps (TOF = $t_{MCP}$ - $t_{Micromegas}$) in the CERN muon test beam. This was obtained using the Cividec amplifier (1-2 GHz BW) and the SAMPIC waveform digitizer, or the LeCroy scope (20 GSa/s). They also achieved $\sigma_{TOF}$ ~36 ps with a laser [33] for Npe ~50.

Proponents hope that this device could be used for solving a calorimeter pile-up problem at the LHC.

(a)

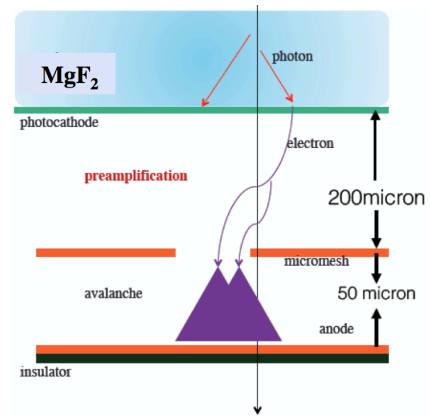

**(b)**

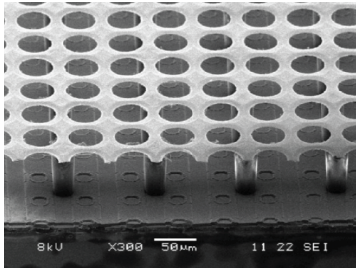

**(c)**

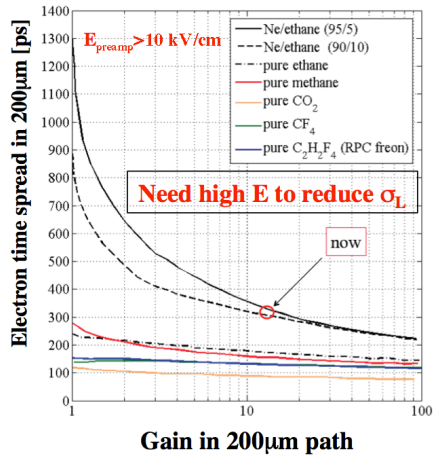

**(d)**

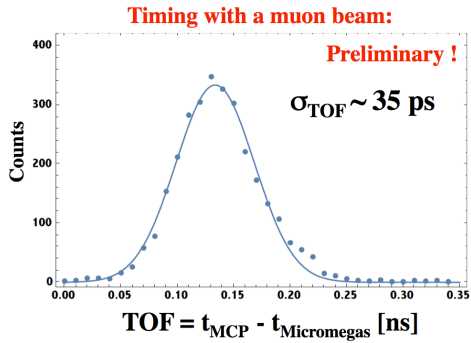

**Fig. 17** (a) Concept of the gaseous detector. (b) Detail of gain region of the Micromegas. (c) Electron time spread per 200 μm drift length as a function of a gain for various gas choices. (d) Excellent timing resolution measured in the CERN muon test beam [28,32,33].

## CONCLUSIONS

- MCP-PMT detectors achieved the best TOF resolution at a level of 9-10 ps, using special single-pixel devices.
- There is a large effort, driven mainly by the LHC, to develop new fast detectors: (a) measuring "position, time and energy," (b) reducing pile-up and (c) developing a Si-tracker providing time resolution at the 20-30 ps level per track.
- Single-pixel MGAD and multi-layer LGAD Si-detectors may achieve 20-30 ps resolution.
- Multi-layer Diamond detectors can also achieve comparable resolution.
- A Si PIN diode with zero gain can be used for high resolution timing with a ~6$X_0$ absorber.
- Gaseous detectors, based on the Micromegas concept, may also be useful for timing in the 30-50 ps domain.
- However, most results in this paper were obtained with single-pixel devices. The timing resolution will be worse in large multi-pixel detector systems.
- The next few years will show what the true outcome will be.

## ACKNOWLEDGEMENTS

I would like to thank A. Rozhnin, S. White, H. Sadrozinski and K. Inami for contributions to this talk.

## REFERENCES

[1] J. Va'vra, RICH 2010, Nucl.Instr. & Meth. A 639 (2011) 193.
[2] N. Harnew, RICH2013, Nucl.Instr. & Meth. A 766 (2014) 274.
[3] dE/dx: H. Bichsel, Nucl.Instr. & Meth. A 562, (2006) 154; and
    PID: C. Lippmann, Nucl.Instr. & Meth. A 666 (2012) 148; and
    Early dE/dx: J. Va'vra, Nucl.Instr. & Meth. A 453 (2000) 262;
    TRD: B. Dolgoshein, Nucl.Instr. & Meth. A 326 (1993) 434;
    TRD: A. Andronic and J.P. Wessels, ArXiv:1111.4188v1 (2011).
[4] Very early TOF; B. Atwood, SLAC-PUB-1980.
[5] ALICE collaboration, J. Adam et al., arXiv: 1602.01392 (2016).
[6] A. Akindinov et al., Eur. Phys. J. Plus, 128 (2013) 44.
[7] J.Va'vra., MCP-PMT log book #1, p.18, 2005, unpublished.
[8] J. Va'vra et al., Nucl.Instr. & Meth. A 606 (2009) 404;
    MCP log book #6, page 122, 2008, unpublished.
[9] N.Kishimoto et al., Nucl.Instr. & Meth. A 564 (2006) 204.
[10] J. Va'vra et al., SLAC-PUB-11934, 2009;
    MCP log book #3, pages 27-40, 2007, unpublished.
[11] J.Va'vra, MCP-PMT log book #4, 2007, unpublished;
    A. Ronzhin et al., Nucl. Instr. & Meth. A 623 (2010) 931.
[12] S. Ritt, http://dx.doi.org/10.1109/TNS.2014.2366071.
[13] A. Ronzhin, Fermilab, talk at SLAC, 2016;
    A. Ronzhin et al., Nucl. Instr. & Meth. A 795 (2015) 288.
[14] K. Inami et al., Nucl. Instr. & Meth., A 560 (2006) 303.
[15] S. White, RD-51 talk, Dec. 4, 2016, private communication.
[16] J. Va'vra et al., Nucl.Instr. & Meth. A 606 (2009) 404.
[17] J. Lange et al., ArXiv:1608.01485 (2016).
[18] A. Brandt, Arlington DOE review workshop, Oct. ,2015.
[19] A. Ronzhin et al., Nucl. Instr. & Meth. A 795 (2015) 52.
[20] H. Frisch, private communication.
[21] J. Wang et al., IEEE Trans. on Nucl. Sci, Nov.23, 2015.
[22] D. Renker and E. Lorenz, "Advances in solid state photon detectors", JINST, April 2009.
[23] A. Fukusawa et al., IEEE at San Diego, 2006.
[24] M. Suyama et al., KEK preprint 2003-134.
[25] A.Ronzhin et al., Nucl.Instr. & Meth. A 623 (2010) 931.
[26] L. Gruber et al., Nucl. Instr. & Meth. A 824 (2016) 104.
[27] M. McClish et al., IEEE TNS, 53, 3049 (2006).
[28] S. White, ArXiv:1409.1165v2 (2014); private communication.
[29] D. Breton et al., Nucl.Instr. & Meth. A 835 (2016) 51.
[30] N. Cartiglia et al., ArXiv:1608.08681v2 (2016).
[31] M. Berrettiet et al., Nucl.Instr. & Meth. A824 (2016) 87;
    J. Pietrasko et al., Nucl.Instr. & Meth. A 618 (2010) 121;
    M. Ciobanu et al., IEEE Trans. on Nucl. Sci, vol. 58, 2011.
[32] Y. Giomataris et al., Nucl.Instr. & Meth. A 560 (2006) 405.
[33] T. Papaevangelou et al., ArXiv:1601.00123v2 (2016);